# Noncontact dielectric constant metrology of low-*k* interconnect films using a near-field scanned microwave probe

Vladimir V. Talanov,[a] André Scherz, Robert L. Moreland, and Andrew R. Schwartz

*Neocera, Inc., 10000 Virginia Manor Road, Beltsville, MD 20705*

We present a method for noncontact, noninvasive measurements of dielectric constant, *k*, of 100-nm- to 1.5-μm-thick blanket low-*k* interconnect films on up to 300 mm in diameter wafers. The method has about 10 micron sampling spot size, and provides <0.3% precision and ±2% accuracy for *k*-value. It is based on a microfabricated near-field scanned microwave probe formed by a 4 GHz parallel strip transmission line resonator tapered down to a few-micron tip size.

[a] Electronic mail: talanov@neocera.com

Operating performance of today's advanced integrated circuits (ICs) is often dominated by interconnect – the multilayer structure consisting of up to ten levels of metal (e.g. Cu) wiring separated by interlevel dielectric (ILD) that distributes clock and other signals and provides power/ground for the various systems in the chip. To reduce the delay times, crosstalk and power consumption in this wiring system, low dielectric constant ILDs with $k$<4 are being implemented to replace $SiO_2$ (the microelectronics community had adopted $k$ for the relative dielectric constant in contrast to $\varepsilon_r$ used in the scientific community) [1]. However, these new 'low-$k$' materials pose significant integration and characterization challenges, such as low mechanical strength and stability, and sensitivity of the dielectric constant to various processing steps [2]. Hence, electrical evaluation of low-$k$ dielectrics in both process development and IC production is a key issue [1].

There are three area-capacitor methods for electrical characterization of *blanket* low-$k$ films, where one electrode is placed directly on the film surface, while the substrate (e.g., Si wafer) serves as a second electrode. Mercury probe provides a quick estimate of $k$-value, however its accuracy is questionable due to interaction between Hg and porous dielectrics [1]; probe-sample cross-contamination is an issue as well. Metal-Insulator-Silicon capacitor is accurate down to 2%, but destructive and time consuming (requires electrode patterning) [1]. The Corona discharge/Kelvin probe method that had been traditionally employed to characterize gate oxides [3] is now being applied to low-$k$ films as well [2], but there is some concern that the charge deposition may cause sample damage. While noncontact, the method is *invasive* since a water rinse is needed to remove the charge from the wafer. None of the above methods is appropriate for porous low-$k$ materials. Thus, rapid, noncontact techniques with smaller spot size are desired for dielectric constant metrology of blanket low-$k$ films. Such a method based on our near-field scanned microwave probe (NSMP) [4] is demonstrated here.

In the past decade NSMPs have emerged as an approach to non-destructive imaging of materials' electrodynamic properties on sub-micron length-scales [5-10]. However, apart from our preliminary study [11] there has been only a single report of dielectric constant measurements of low-$k$ films with an evanescent microwave probe [12]. This work utilized a 100-micron tip in physical contact with the sample and for a variety of blanket low-$k$ samples the probe resonant frequency shift was shown to scale with the film thickness divided by $k$-value. No precision and accuracy for the extracted $k$-values were reported.

We review some basic shortcomings of the existing NSMPs that have likely limited their use in the study of low-$k$ dielectrics. First, until recently [4, 8, 13] most probes have been lacking an active tip-sample distance control mechanism, which is crucial since the near-field probe response is extremely sensitive to this "lift-off" distance [14]. Second, since the *near-zone* field distribution depends strongly on the antenna geometry [15], even a slight but inevitable imperfection in the probe shape and/or alignment with respect to the sample substantially changes the interaction with the sample. This makes it virtually impossible to rely *entirely* on calculations to accurately deduce the sample permittivity from measured results. In addition, only the simplest geometries, such as a metallic sphere [16, 17] or a flush open end of a coaxial transmission line [14], can be solved analytically. A semi-empirical approach, on the other hand, employs standard samples with known properties to create a calibration data set and find the properties of samples under test by converting the measured quantity into the unknown value via some interpolation scheme [5–8, 11, 12]. For thin (e.g., few hundred nanometers thick) dielectric films, however, this is complicated by the lack of established standards: the only practically available films with accurately known permittivity are air ($k$=1.00), thermally grown $SiO_2$ ($k$=3.93±0.01), and metal ($|k|\gg 1$). To interpolate the entire calibration curve from these points, an analytical or numerical model is needed, which typically requires that the probe geometry and *absolute* tip-sample distance be known with impractically high accuracy. The calibration approach developed here overcomes these difficulties.

Our probe is a microfabricated $\lambda/2$ parallel strip transmission line resonator (PSR), which is formed by a $SiO_2$ micropipette with square cross-section pulled down to a few micron tip size and sandwiched between two aluminum strips of thickness $a$~3 μm [4]. The PSR is packaged inside a metallic sheath with the taper protruding out via an opening in the sheath wall (Fig.1). It operates in the balanced odd mode with a resonant frequency ~4 GHz. An electrical near-field similar to the fringe field of a parallel plate capacitor is formed at the probe tip. When a dielectric sample is brought in close proximity to the tip, the reactive energy stored in this field is reduced, and, consequently, the probe resonant frequency $F$ decreases. Since the tip size $D$~1–10 μm is much smaller than the radiation wavelength, a lumped element scheme can be used. For a thin low-loss film on an arbitrary substrate the tip impedance $Z_t$ terminating the PSR can be represented as a network of the air-gap capacitance $C_g$, the film capacitance $C_f$, and the substrate impedance $Z_s$ (Fig.1). For a bulk substrate with thickness $\gg D$ and complex relative permittivity $k_s$ one can estimate $Z_s \approx 1/i\omega\varepsilon_0 k_s D$ [10]. If $|Z_s| \ll 2/\omega C_f$ its contribution is negligible with respect to other capacitive impedances. Assuming a parallel plate model for $C_f$ this condition can be met if $|k_s| \gg ka/2t_f$, which yields $|k_s| \gg 15$ at 4 GHz for typical parameters ($k \sim 3$, $t_f \sim 300$ nm, $a \sim 3$ μm). This can be achieved, for example, on a Si substrate with resistivity <0.3 Ω·cm or on a metal substrate. Finally, $Z_t = 2(C_f^{-1} + C_g^{-1})/i\omega = 1/i\omega C_t$ and the relative shift of the probe resonant frequency, $\Delta F$, versus change in the tip capacitance $C_t$ is [11]:

$$\Delta F/F = -2FZ_0\Delta C_t \tag{1}$$

where $Z_0 \approx 100\,\Omega$ is the transmission line characteristic impedance, and $FZ_0 C_t \ll 1$. Eq. (1) predicts the capacitance sensitivity for our probe ~0.3 aF where $F$ is determined with a precision ~1 kHz by a frequency counter measuring the carrier frequency of a voltage controlled oscillator locked onto the resonance.

The data acquired by our NSMP is the dependence of $\Delta F(h) = F_0 - F(h)$ on the tip-sample distance $h$ that is varied with sub-nanometer precision using a piezo stage. Here $F_0$ is the probe frequency without sample present. The closest distance, $h_{SF}$, to which the tip is brought over a sample is controlled with ~2 nm precision using a shear-force (SF) method with optical detection [4]. Assuming the tip end is flat and parallel to the sample surface, $h_{SF}$ is estimated to be ~50–100 nm based on typical tip geometry and resonant frequency shift $F_0 - F(h_{SF}) \sim 1$ MHz. We speculate, however, that due to inevitable imperfections in the real tip shape and alignment with respect to the sample there is a point on the tip that is much closer to the sample (e.g. ~10 nm), which is actually responsible for the SF interaction. Wafers up to 300 mm can be scanned with a 350×350 mm travel xy-stage beneath the probe. The apparatus sits on a vibration-isolated platform at ambient conditions inside an environmental chamber.

Dependence of the probe frequency $F$ on film dielectric constant $k$ is governed by the probe geometry, $h$, $t_f$, and $Z_s$. The $Z_s$ contribution was kept negligibly small for all films, including calibration films, by fabricating them on low resistivity (<5 mΩ·cm) silicon wafers. $F$ is a monotonically decreasing function of $k$ and of the parameter $\beta = (k-1)/(k+1)$ pertinent to any electrostatic problem involving a charged metallic body placed near a bulk or layered dielectric [16]. As $k$ varies from 1 to infinity, $\beta$ ranges from 0 to 1. We have experimentally observed and confirmed by 2D finite element electrostatic modeling that at a given film thickness the *curvature* of $\Delta F(\beta)$ or $\Delta C_t(\beta)$ monotonically depends on $h$ such that at a certain $h = h^*$, $\Delta F(\beta)$ becomes *linear* for all $\beta$. This is shown in Fig. 2a, where it is also clear that the $\Delta F(\beta)$ curvature is positive for $h<h^*$ and negative for $h>h^*$. This behavior can be qualitatively explained assuming that $C_f=kC_{f0}$ and $C_g=C_{g0}$ (Fig. 1), where $C_{f0}$ and $C_{g0}$ are the geometrical capacitances. If $C_{f0}=C_{g0}$ then $\Delta F \propto \Delta C_t$ is a linear function of $\beta$. Two parallel plate capacitors of the same area have the same geometrical capacitances if their thicknesses are equal, hence in this simple model $\Delta F(\beta)$ is linear when $h^*=t_f$. Therefore, if the measurements are made at the unique tip-sample distance $h^*$ only two "standard" films with $\beta=0$ (e.g., air) and $\beta=1$ (e.g., metal) are needed to calibrate the probe throughout the entire $\beta$ range. Clearly, any bulk conductor with low enough resistivity (we use Si with $\rho<5$ m$\Omega$·cm) or any metallic film with sheet resistance $R_{sh}<<1/\omega C_g$ can be used in place of the film with $\beta=1$.

In order to find the dependence of $h^*$ on $t_f$, which is unique to the probe geometry, we employ a set of six thermally grown SiO$_2$ films with variable thickness ranging from 0.1 to 1.5 µm. Figure 2b shows the typical measured dependence along with the parallel plate capacitor model and 2D simulation results. The measured and simulated dependencies differ because of *inevitable* uncertainty in the tip geometry and the 2D rather than 3D simulation. Unlike the parallel plate model, they both yield a slope <1 due to fringing effects (the capacitors mentioned above are not perfect parallel plates). Extrapolating the experimental $h^*(t_f)$ to $t_f=0$ yields $h_{SF}$~80 nm, which is in agreement with the estimate mentioned above. In the simulations we also used $h_{SF}=80$ nm. It is evident that $h^*(t_f)$ is a monotonic function close to linear and can be easily interpolated to determine $h^*$ for arbitrary film thickness. Note, that at $h=h^*$ the probe is still sensitive to the film permittivity since $C_f/C_g$~$kh^*/t_f$~1 for $h^*\leq t_f$.

Once $h^*$ vs. $t_f$ is established, the quantitative measurement of a film under test is straightforward. First, $h^*$ is determined for the thickness of the film under test, $t_{fut}$. Second, $\Delta F$ is measured at $h=h^*(t_{fut})$ on two "standard" $t_{fut}$-thick films with $\beta=0$ and $\beta=1$, and a *linear* calibration curve $\Delta F(\beta)$ is constructed. Third, to determine the dielectric constant of the film under test, the frequency shift is measured at the same tip-sample distance $h^*(t_{fut})$ and converted into a $k$-value via the linear $\Delta F(\beta)$ calibration curve. The important assumption made is that the shear-force height $h_{SF}$ is nominally the same for all films. Our experience shows that this holds well for samples with surface roughness up to at least a few nanometers.

To validate our ability for quantitative measurements we investigated more than twenty blanket films on 200 and 300 mm Si wafers. The set of films contained various spin-on and CVD low-$k$ dielectrics, porous and non-porous, organic and inorganic, with thicknesses ranging from 0.1 to 1.5 μm and $k$-values between 2 and 3.5, as well as a few high-$k$ films. Figure 3a shows excellent correlation between our data and the results obtained by Hg-probe at 0.1 or 1 MHz with a correlation coefficient $R^2$= 0.998. The short-term repeatability of our technique has a 1σ<0.3% (Fig. 3b). It is mostly limited by repeatability of the tip-sample distance control rather than the frequency measurement. Since no model describing probe-sample interaction is involved in converting the measured quantities, frequency and distance, into the film dielectric constant, the accuracy is mostly due to interpolation of the $h^*(t_f)$ curve, which is estimated to be better than ±2% (provided the film thickness is determined by ellipsometry or reflectometry with an accuracy down to a fraction of a percent). It can be improved even further by using additional $SiO_2$ films within the desired thickness range.

To conclude, we have developed an accurate and precise method for dielectric constant metrology of blanket low-$k$ films with the following advantages: 1) it requires no knowledge of either tip geometry or *absolute* tip-sample distance, and no analytical or numerical modeling for the probe-sample interaction is involved; 2) it can be used for the entire range of film dielectric constants and for film thicknesses down to at least 100 nm; 3) the noncontact, noninvasive nature and small sampling area ~10 μm [4] make the technique particularly well suited for electrical metrology within the scribe line or active die regions on semiconductor device wafers, which will be reported elsewhere. The method can also be used to accurately measure permittivity of a thin film on an arbitrary substrate (either insulating or metallic) characterized by complex permittivity $k_s$ if the condition $|k_s|\gg ka/2t_f$ is satisfied.

This work was partially supported by NSF-SBIR 0078486 and NIST-ATP 70NANB2H3005. We thank Professor T. Venky Venkatesan, J. Matthews, Dr. Y. Liu, and Professor S. Anlage for stimulating discussions.

# Figure captions

**Figure 1.**

Apparatus schematic and the lumped element scheme for the tip impedance in the case of a thin dielectric film on an arbitrary substrate. Tip-sample distance $h\sim50-100$ nm; tip size $D\sim1-10$ μm.

**Figure 2.**

(a) 2D finite element model simulations of the normalized frequency shift $\Delta F(\beta) \propto [C_t(\beta)-C_t(0)]/[C_t(1)-C_t(0)]$ for a 250-nm-thick film on a metallic substrate at three tip-sample distances $h$=50, 225, and 500 nm. The tip geometry is shown in Fig. 1. It is evident that $\Delta F(\beta)$ is linear (within simulation precision <0.1%) at $h^*$=225nm. For these simulations $h_{SF}$=0. Solid lines are guides to the eye.

(b) Typical experimental dependence for $h^*$ vs. $t_f$ obtained on a set of six SiO$_2$ films, the parallel plate capacitor model $h^*=t_f-h_{SF}$, and results of 2D simulation. Solid lines are guides to the eye. Note that in this graph $h^*$ is not absolute but is measured relative to $h_{SF}$=80nm, which was determined by extrapolating the experimental data to $t_f$=0.

**Figure 3.**

(a) Correlation between NSMP and Hg-probe measurements on log-log scale for a number of low-$k$ and high-$k$ films. Solid line is the linear fit with correlation coefficient $R^2$=0.998.

(b) Histogram based on 100 measurements performed at the same site on a ~400-nm-thick low-$k$ film. Mean $k$-value is 3.237, with a standard deviation of 0.29%.

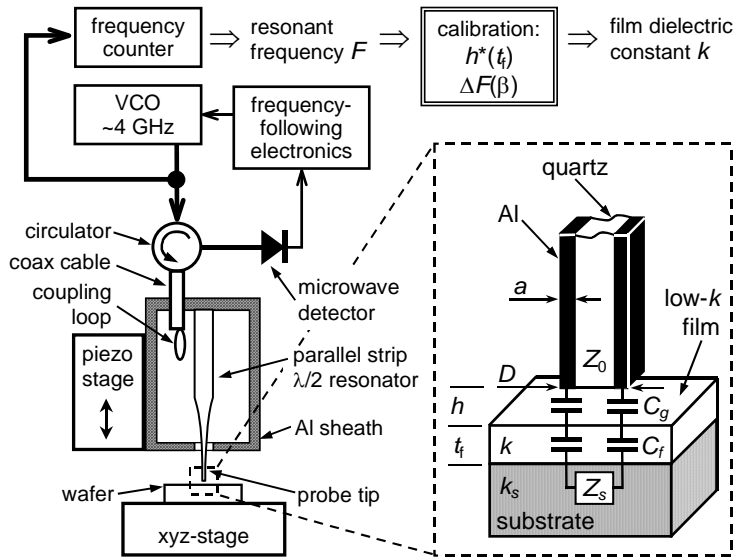

**Fig. 1**

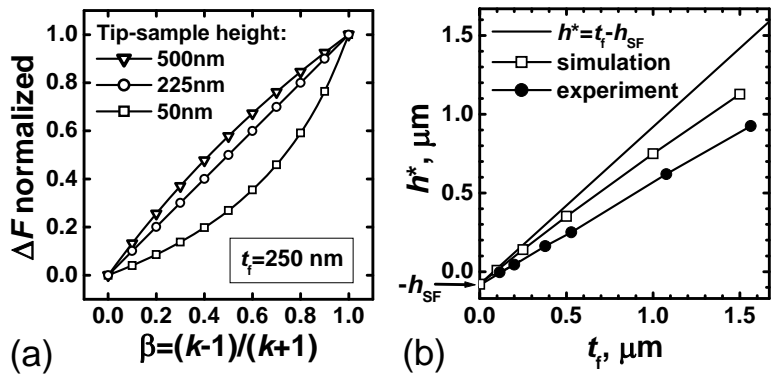

**Fig. 2ab**

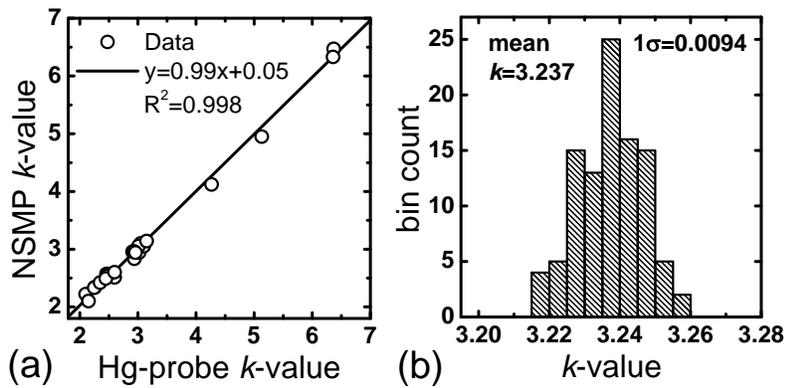

**Fig. 3ab**

9